\documentclass[aip, jcp, reprint, floatfix, endfloat, showkeys]{revtex4-1}

\usepackage{amsmath,amsbsy,amssymb,amsfonts}
\usepackage{graphicx, color}
\usepackage{empheq}
\usepackage{marginnote}
\usepackage{subfigure}
\newcommand{\angstrom}{\textup{\AA}}
\newcommand{\Matr}[1]{\boldsymbol{\mathcal{\hat{#1}}}}

\graphicspath{{Pictures/}}

\begin{document}
\date{\today}
\title{Electronic torsional sound in linear atomic chains: chemical energy transport at 1000~km/s.}
\author{Arkady A. \surname{Kurnosov}}
\author{Igor V. \surname{Rubtsov}}
\author{Andrii O. \surname{Maksymov}}
\author{Alexander L. \surname{Burin}}
\email[Author to whom correspondence should be addressed. Email: ]{aburin@tulane.edu}
\affiliation{Department of Chemistry, Tulane University, New Orleans, Louisiana 70118, USA}

\begin{abstract}
We investigate entirely electronic torsional vibrational modes in linear cumulene chains. The carbon nuclei of a cumulene are positioned  along the primary axis so they can participate only in transverse and longitudinal motions. However,  the interatomic electronic clouds behave as a torsion spring with remarkable torsional stiffness. 
The collective dynamics of these clouds can be described in terms of electronic vibrational quanta, which we name \textit{torsitons}. It is shown that the group velocity of the wavepacket of \textit{torsitons} is much higher than the typical speed of sound, because of the small mass of participating electrons compared to the atomic mass. 
For the same reason the maximum energy of the \textit{torsitons} in cumulenes is as high as a few electronvolts, while the minimum possible energy is evaluated as a few hundred wavenumbers  and this minimum is associated with asymmetry of zero point atomic vibrations. 
Molecular systems for experimental evaluation of the predictions are proposed.     
\end{abstract}

\maketitle


\section{\label{Sec:Into}Introduction}
Highly efficient and fast vibrational energy transport on a molecular scale has been a subject of theoretical and experimental investigations in recent decades. The possible applications in biochemistry, organic chemistry and nanotechnology include development of efficient cooling in microscopic and nanoscopic molecular systems, such   as  nanowires \cite{segal2003thermal}    and   optical   limiters, designing efficient  energy  transport  schematics  for  energy  signaling \cite{lin2012constant}, as well as optimizing  and  even promoting  chemical  reactions  by concentrating the excess energy at the reaction center \cite{lin2009modulating, glowacki2011ultrafast}. It is suggested that quantum vibrational excitations can be manipulated similarly to electrons and photons, thus  enabling  controlled  heat  transport.  Moreover, delocalized excitations (phonons) can be used to carry and process quantum information \cite{gruebele2004vibrational, li2012colloquium, weidinger2007quantum}. The highest transport speed was found in alkanes (1.44 km/s) \cite{yue2015band}.

Possible candidates capable to maintain fast and efficient energy transport are oligomers because of their periodic structure \cite{rubtsova2015ballistic}. In such systems  vibrational states can be substantially delocalized because of the  strong interaction of equivalent site states, so that ballistic energy transport takes place as a free-propagating wavepacket. The ballistic constant-speed transport has been observed in bridged azulene-anthracene compounds \cite{schwarzer2004intramolecular}, polyethylene glycol oligomers \cite{lin2012constant}, alkanes \cite{rubtsova2015ballistic, wang2007ultrafast, yue2015band}, and   perfluoroalkanes \cite{rubtsova2013ballistic, rubtsova2014temperature} and the theory describing this transport and its possible breakdown due to decoherence has been suggested \cite{segal2003thermal, benderskii2011propagating, benderskii2013vibrational, kurnosov2015communication}. 

Phonon wavepackets in carbon based polymer chains can propagate with the group velocity as high as 10~km/s because of a high strength of covalent bonds  \cite{henry2008high, shen2010polyethylene}. Yet the maximum energy of singly excited vibrations does not exceed ca. 3000~cm$^{-1}$, as the motion is associated with displacements of rather heavy nuclei. Thus, the ballistically transferred energy is much smaller than a typical bond energy exceeding 1~eV ($\sim10^5$~cm$^{-1}$).
 Involvement of multi-phonon transport to increase the amount of transferred energy is expected to enhance the energy relaxation/dissipation. Much larger energy can be carried by excitons, delocalized electronic states \cite{davydov1962theory}. However molecular excitons are usually strongly coupled to the environment resulting in incoherent energy transport (see e. g. exciton transport in DNA \cite{lewis2005dna, burin2009sum}).  

\begin{figure}
\centering
\subfigure[cumelene]{\label{Fig:C5H4}\includegraphics[scale = 0.3]{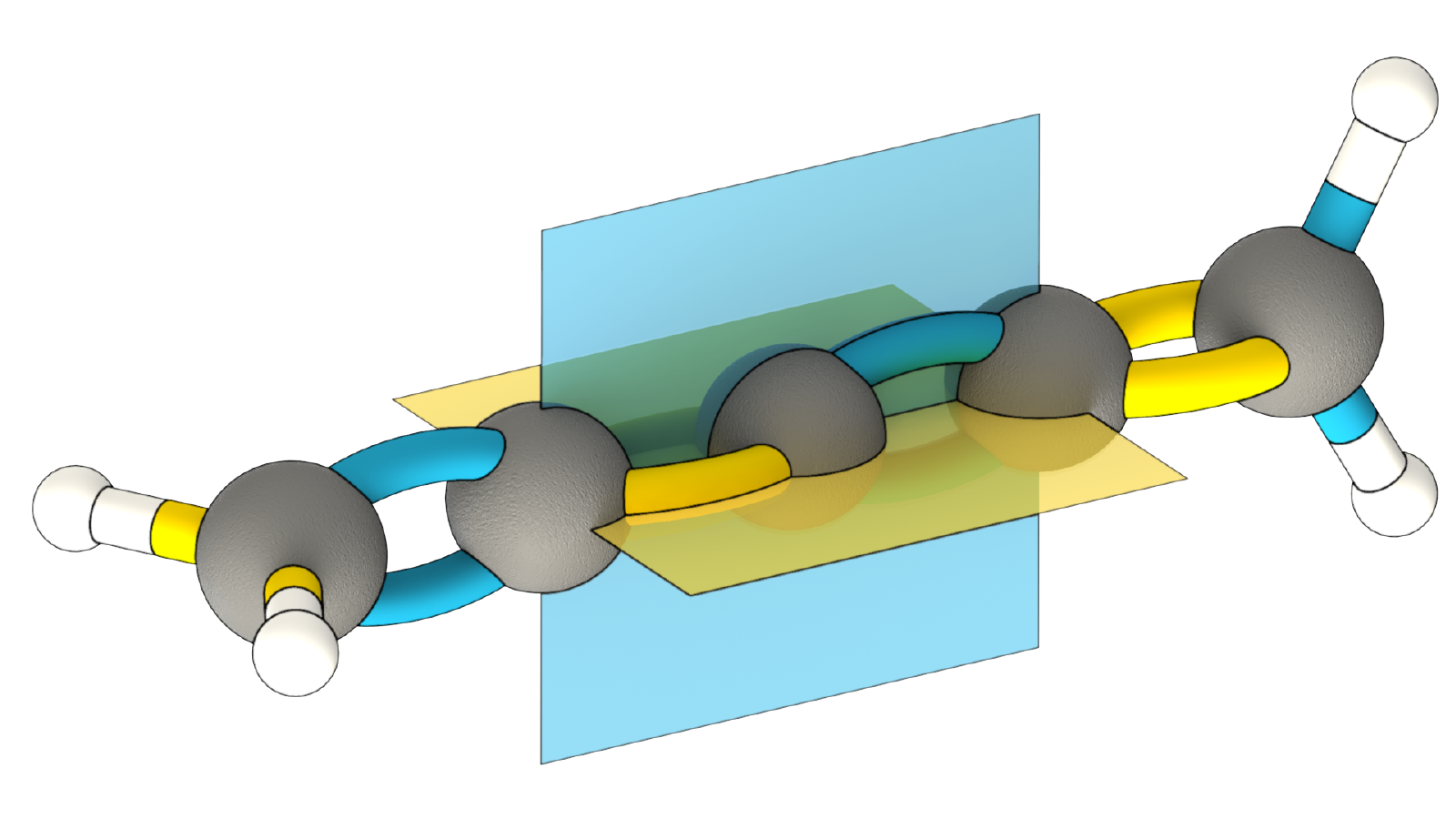}}\\
\subfigure[polyyne]{\label{Fig:C4H2}\includegraphics[scale = 0.3]{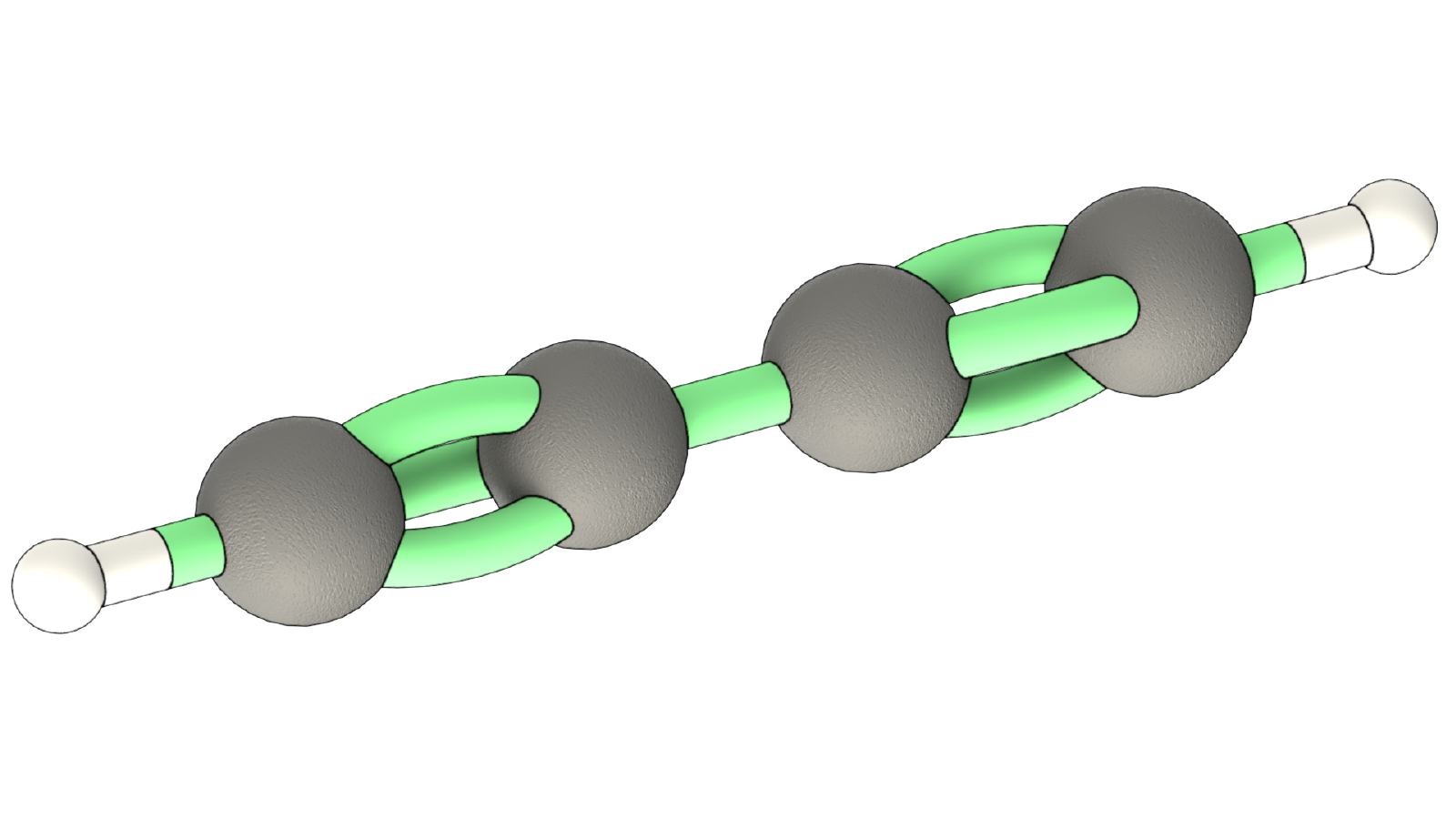}}
\caption{\label{Fig:Carbynes} Carbyne modifications: (a) cumulene molecule with orthogonal double bonds; (b) polyyne molecule with alternating single and triple bonds}
\end{figure}

Here we propose to exploit the special vibrational modes of entirely electronic nature capable of efficient delivering energies in the eV range. Such modes can exist in molecules having all atoms aligned along the single axis (see Figure~\ref{Fig:Carbynes}) and they are formed by propagating torsional  oscillations of  electronic nature. Nuclei do not participate in these oscillations because their rotation about the axis they located on is degenerate.

\begin{figure}
\centering
\includegraphics[scale = 0.3]{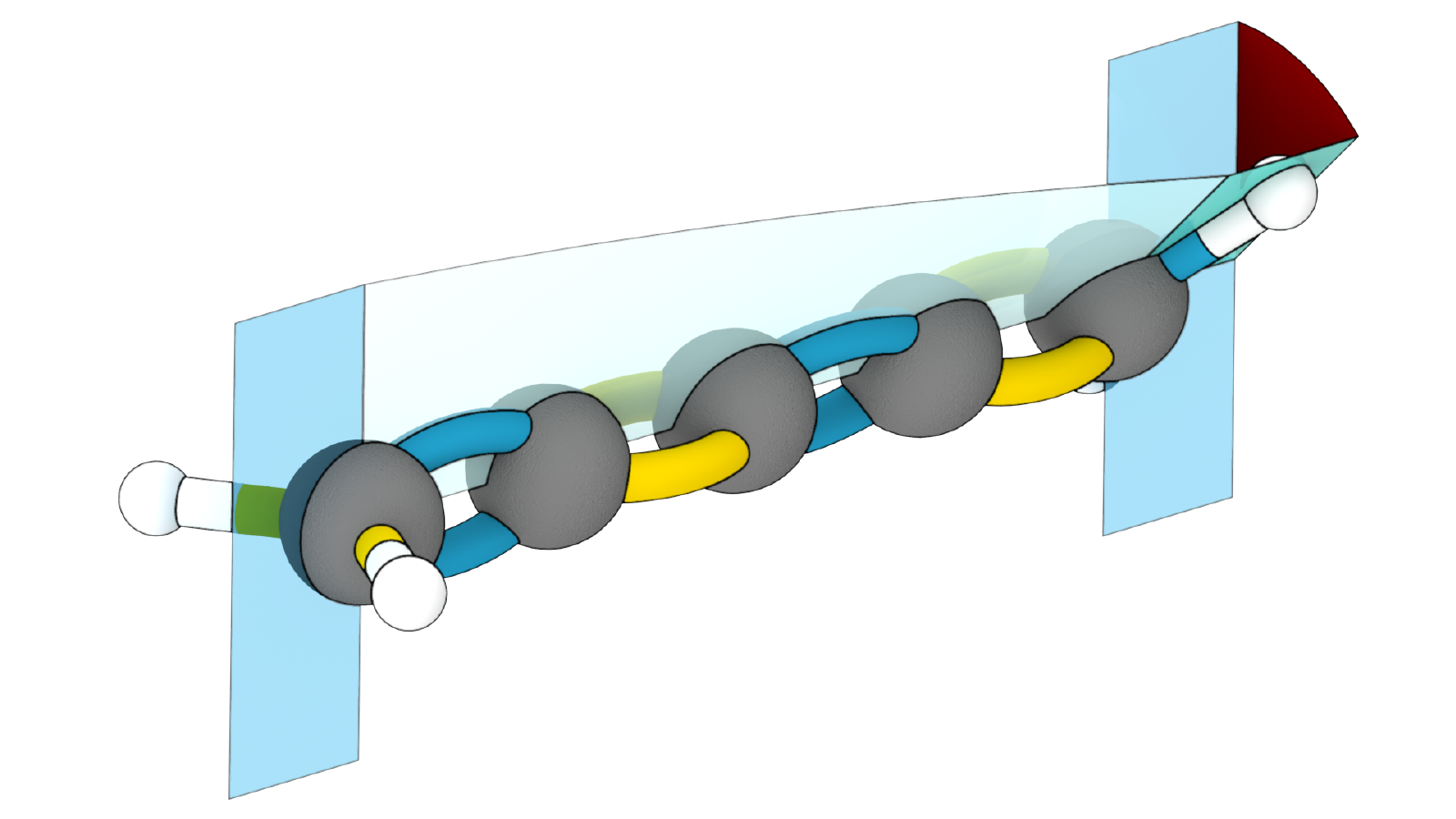}
\caption{\label{Fig:C5H4Twist} Cumulene molecule torsionally strained along primary axis}
\end{figure}

Considering a linear molecule as an elastic rod,  four gapless phonon branches are expected based on symmetry, including longitudinal, two transverse, and one torsional modes \cite{piseri1968dispersion, jishi1993phonon}. The longitudinal and torsional oscillations of frequency $\omega$ and wavevector $q$ are characterized by an acoustic spectrum $\omega = cq$ with a relatively high speed of sound, $c$. As oppose to an elastic rod, for a molecular chain with all atoms located on the same axis, there is no nuclear contribution to the torsional vibrations, since the chain is completely linear.
 Nevertheless, the system can possess a remarkable torsional stiffness due to anisotropic arrangement of its electronic clouds. Such situation is found in cumulenes, featuring a chain of carbon atoms coupled to each other by double bonds \cite{liu2013carbyne}, where the anisotropy results from the $\pi$-bond anisotropy between carbon atoms (see Figures~\ref{Fig:C5H4}, \ref{Fig:C5H4Twist}).
Similar conditions can be realized in transition metals where atoms can form chain bridges between junctions \cite{hasmy2008formation, dai2011metal, tereshko2013biomedical}. 

The torsional sound should exist in such system and we expect it to be of a purely electronic nature because nuclei are positioned along the primary axis and cannot participate in the torsional motion. Since electrons are much lighter than atoms it is natural to expect the speed and a single quantum energy to be much higher than those for nuclei vibrations. 

In the present study we performed a first principle investigation of  the electronic torsional waves in cumulene chains.
We found that the speed of sound in cumulenes to be as high as 1000~km/s and a maximum energy of the quantum as high as 10~eV. 
Because this type of motion is not related to the atomic vibrations and differs from Langmuir waves in plasma, the corresponding quantum quasi-particle is neither a phonon nor a plasmon. To avoid confusion we will call a quantum of torsional electronic oscillations a \textit{torsiton}. The effect of zero point vibrations leading to the \textit{torsiton} spectral gap of the order of 0.03~eV is estimated. The possible ways to observe \textit{torsitons} experimentally are discussed.

We consider the torsional oscillations of a cumulene in a dielectric environment, so the electronic excitations can be neglected. Although the metallic behavior of cumulene was predicted theoretically \cite{liu2013carbyne, larade2001conductance} it is not confirmed experimentally \cite{cretu2013electrical, banhart2015chains} so the nature of electronic excitations remains unclear. Here we ignore electronic excitations assuming that there is the significant spectral gap (cf. Ref. \cite{cretu2013electrical}).

\section{\label{Sec:System}The system}
A linear cumulene chain is a compound containing a sequence of $n$ carbon atoms with $(n - 1)$ double bonds between them R=C=(C=)$_{n-2}$C=R \cite{fischer1964chemistry}.
Quantum chemistry calculations were performed for the simplest termination of cumulene chain by two hydrogen atoms on each side; an example of cumulene molecule  H$_2$C=(C=)$_3$CH$_2$ is shown schematically in Figure~\ref{Fig:C5H4}. One can see that orthogonal $\pi$-bonds between carbon atoms can provide rigidity with respect to twisting with remarkable torsional stiffness, while much smaller stiffness is expected in another carbyne modification, polyyne, which is a chain of carbon atoms with alternating single and triple bonds between them (see Figure~\ref{Fig:C4H2}). For cumulene molecules the shortened notation H$_2$C$_n$H$_2$ (without bond type specification) will be used.


\section{\label{Sec:Speed}Electronic torsional mode}
To estimate the speed of sound for electronic torsional wave we consider a model of elastic rod (torsion spring) which can be described by the Lagrangian 
\begin{equation}\label{Eq:RodLagr}
\mathcal{L}_{e} = \frac{1}{2}\int\limits_{z_l}^{z_r}dz \Bigg\{j_e\left(\frac{d\theta}{d t}\right)^2 - \kappa\left(\frac{\partial\theta}{\partial z}\right)^2\Bigg\}
\end{equation}
where dynamical variable $\theta(z, t)$ is a twisting angle of the rod along z-axis as a function of  coordinate along prime axis and time; two neighboring cross-sections at points $z$ and $z + dz$ will rotate with respect to each other with a relative angle $d\theta = \left(\partial\theta/\partial z\right)dz$  \cite{landafshitz2}.  
Here $z_{l, r} = \mp L/2$, $L$ - molecule length, $\kappa$ stands for torsional stiffness and $j_e$ is an average linear density of electronic moment of inertia with respect to z-axis. 

We estimate parameters of interest as $j_e = 1.73$~$m_e\cdot\angstrom$ ($0.95\cdot 10^{-3}$~u$\cdot\angstrom$) and torsional stiffness  $\kappa = 10.6$~eV$\cdot\angstrom$ as described in the next two sections. Our estimate for the torsional stiffness is consistent with the previous estimate of 10.3~eV$\cdot\angstrom$ reported in Ref. \cite{liu2013carbyne}.

With the angle $\theta(z, t)$ and the related angular velocity $d\theta/dt$ considered as dynamical variables Eq. (\ref{Eq:RodLagr}) leads to the Euler equation  
\begin{equation}\label{Eq:FastModeMotion}
j_e\frac{\partial^2\theta}{\partial t^2} = \kappa\frac{\partial^2\theta}{\partial z^2}
\end{equation}
which is a wave equation with the dispersion relation $\omega(q) = q\sqrt{\kappa/j_e}$ and the speed of \textit{torsiton} wave
\begin{equation}\label{Eq:SpeedOfSound}
c = \frac{d\omega}{d q} = \sqrt{\frac{\kappa}{j_e}} \simeq 1.0\cdot 10^6 \, \text{m/s}
\end{equation}
using $j_e$ evaluated below.
This velocity exceeds the typical phonon propagation velocity in polymers by two or three orders of magnitude. 
Next we also estimate the maximum energy transfered by the electronic torsional mode. 

The dispersion relation for longitudinal vibration in a uniform chain with nearest neighbor coupling and the lattice period $a$  has the standard form $\omega(q) = \omega_*\sin(aq/2)$ \cite{nitzan2006chemical}.

It should be a good approximation for the torsional mode under consideration because the interaction responsible for the torsional stiffness is due to short range covalent bonding. In the long wavelength limit $q\longrightarrow 0$ we estimate maximum energy of the \textit{torsiton} as
\begin{equation}\label{Eq:energy} 
\hbar\omega_* = 2\frac{\hbar c}{a}\simeq 10.5 \, \text{eV} 
\end{equation}
(the lattice period in cumulenes is given by $a = 1.28$~$\angstrom$ \cite{cahangirov2010long}).  This result corresponds to the nearest neighbor effective coupling $\hbar\omega_*/4 \sim 2.6$~eV, which is  about the $\pi$-bond strength of 2.74~eV in  C=C bonds.

We can also roughly estimate the \textit{torsiton} mean free path order of magnitude.
In Ref. \cite{rubtsova2014temperature} the decoherence rate has been estimated analyzing the switch between ballistic and diffusive transport in perflouroalkanes. It is found to be $W = 2$~ps$^{-1}$, so the phonon decoherence time has been estimated as $T\sim 1/W\sim 0.5$~ps. Assuming similar decoherence time for the \textit{torsiton}, one can estimate its mean free path as $l_0 = c T \sim 0.5$~$\mu$m.   

Thus we found the electronic torsional sound wave velocity and energy unprecedentedly high compared to typical phonon parameters which makes this system very attractive for energy transport applications. The energy transferred by a single quantum is sufficiently large for chemical applications: bond making-bond breaking, energy release, and energy transfer to reaction center.  

Below we derive our estimate for electronic moment of inertia and for torsional stiffness, discuss the limitations of our result due to zero point atomic vibrations and propose the way to observe the ultrafast energy transport due to electronic torsional sound.  

\section{\label{Sec:je}Electronic moment of inertia}
The linear density of electronic moment of inertia is defined as 
\begin{equation}\label{Eq:InMomLinDens}
j_e(z) = \iint\rho(x, y, z)\left(x^2 + y^2\right)dxdy  
\end{equation}
where $\rho$ is the electronic density.
We calculated the linear density of electronic moment of inertia for cumulene molecule using density functional theory with \texttt{B3LYP} hybrid functional and standard \texttt{6-31(d, p)} basis set, as implemented in a Gaussian~09 software package \cite{g09}. The electron density as a function of coordinates is extracted with the uniform grid of 0.1~Bohr radius (0.0529~\AA), the symmetric limits in X-Y plane (the plane perpendicular to the prime axis) were chosen $\pm 6.5$~Bohr radius ($\pm 3.44$~\AA). Either doubling of the limits or decrease of the grid by the same factor change the result by less than $1\%$.     

To estimate the accuracy of the numerical result we tested the same approach on the hydrogen atom. The theoretical value of the moment of inertia of hydrogen electron cloud in the ground state can be calculated using the electron wave-function \cite{landau2013quantum} as  $J_0 = 2 m_e a_0^2$, where $m_e$ is electron mass and $a_0$ is the Bohr radius. The result of numerical calculations obtained using the same method as for cumulene is $J_{num} = 1.90 m_e a_0^2$, which is within $5\%$ accuracy.  

\begin{figure}
\centering
\includegraphics[scale = 0.4]{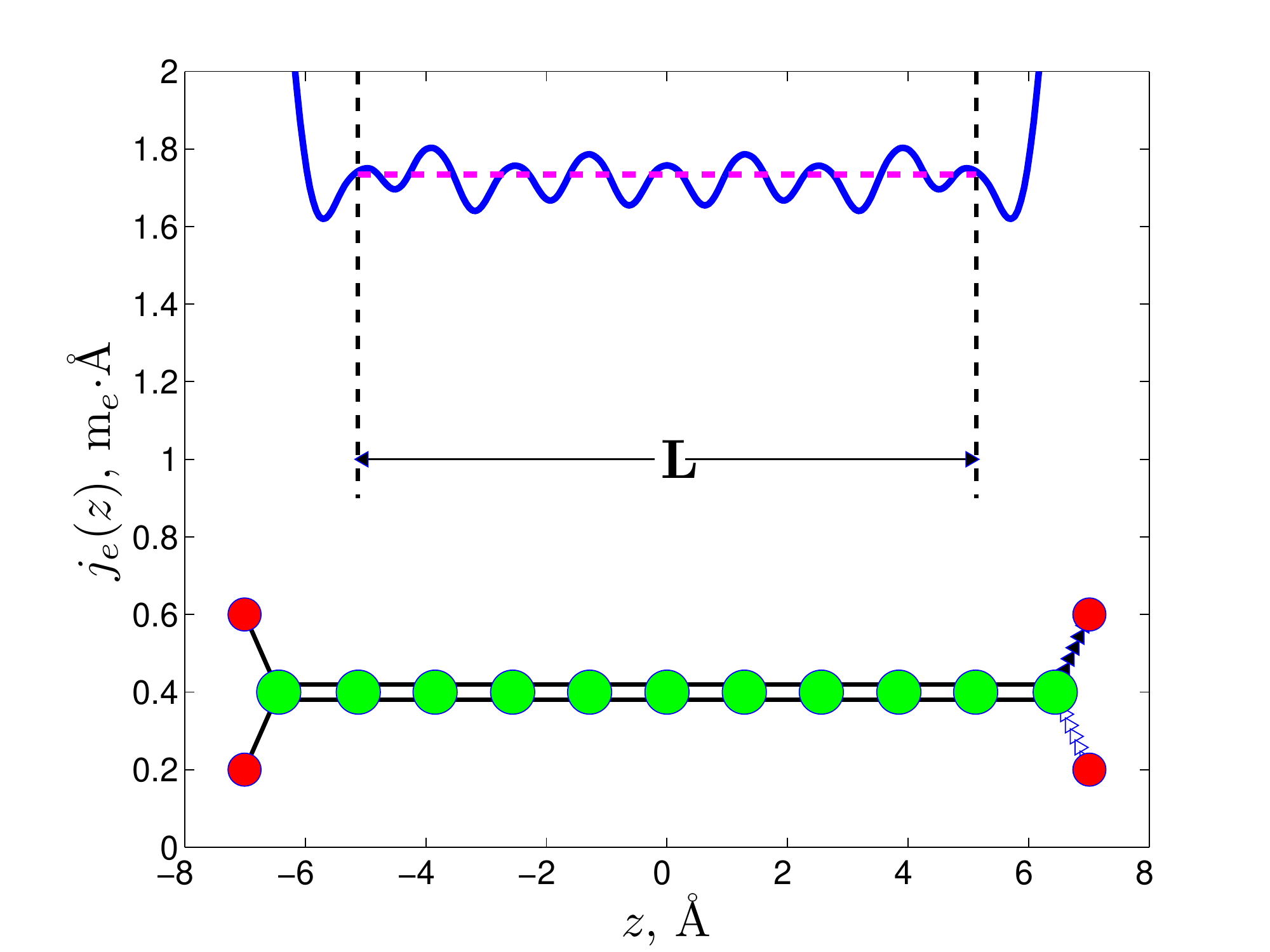}
\caption{\label{Fig:JeVSz}Linear density of the electronic inertia moment $j_e(z)$ as a function of coordinate along primary axis $z$ (blue solid line). The corresponding molecule H$_2$C$_{11}$H$_2$ is represented below. One can see that $j_e(z)$ is a smooth function with low deviations from the average (magenta dash-line), except of the boundaries. The effective length $L$ is defined as a distance between second left and second right carbon atoms.}
\end{figure}

In Figure~\ref{Fig:JeVSz} we show dependence $j_e(z)$ obtained from DFT-calculations for H$_2$C$_{11}$H$_2$. One can see that $j_e(z)$ is a smooth function weakly deviating from its average  $J_e/L$ ($\sim 5\%$), so for simplicity coordinate dependent moment of inertia density $j_e(z)$ can be replaced with the constant $j_e\simeq J_e/L\simeq 1.73$~$m_e\cdot\angstrom$ ($0.95\cdot 10^{-3}$~u$\cdot\angstrom$). As shown in Figure~\ref{Fig:JeVSz}, we define molecular length $L$ as a distance between the second left and second right carbon atoms, where $j_e(z)$ is still not affected by boundaries.

\section{\label{Sec:kappa}Torsional stiffness calculation}
Combining the proposed model with the first principles calculations of the hydrogen atom torsional vibrational mode associated with the relative torsional oscillations of pair of hydrogen atoms (``whiskers", see  Figure~\ref{Fig:C5H4Twist}), we introduce Lagrangian
\begin{equation}\label{Eq:SpringLagr}
\mathcal{L} = \frac{J_l}{2}\left(\frac{d\Phi_l}{d t}\right)^2 + \frac{J_r}{2}\left(\frac{d\Phi_r}{d t}\right)^2 - \int\limits_{-L/2}^{+L/2}dz\frac{\kappa}{2}\left(\frac{\partial\theta}{\partial z}\right)^2
\end{equation}
where $\Phi_l(t)$, $\Phi_r(t)$ are the angles of the ``whiskers" deviation from equilibrium on the left and right side, $J_l = J_r = J/2$ are moments of inertia of the whiskers, $J$ is the entire atomic moment of inertia along primary axis, defined by 4 hydrogen atoms. $\theta(z, t)$ is the same as in Eq. (\ref{Eq:RodLagr}) with boundary conditions $\theta(\mp L/2) = \Phi_{l,r}$. 

In this model the potential energy is originated from the torsional strain of the electronic spring and kinetic energy is entirely defined by the motion of hydrogen atoms, so long as the kinetic energy of electrons is neglected. The latter assumption is justified as long as $J_e \ll J$ (0.028 vs 3.43~u$\cdot\angstrom$ for $n = 25$).  

\begin{figure}
\centering
\includegraphics[scale = 0.4]{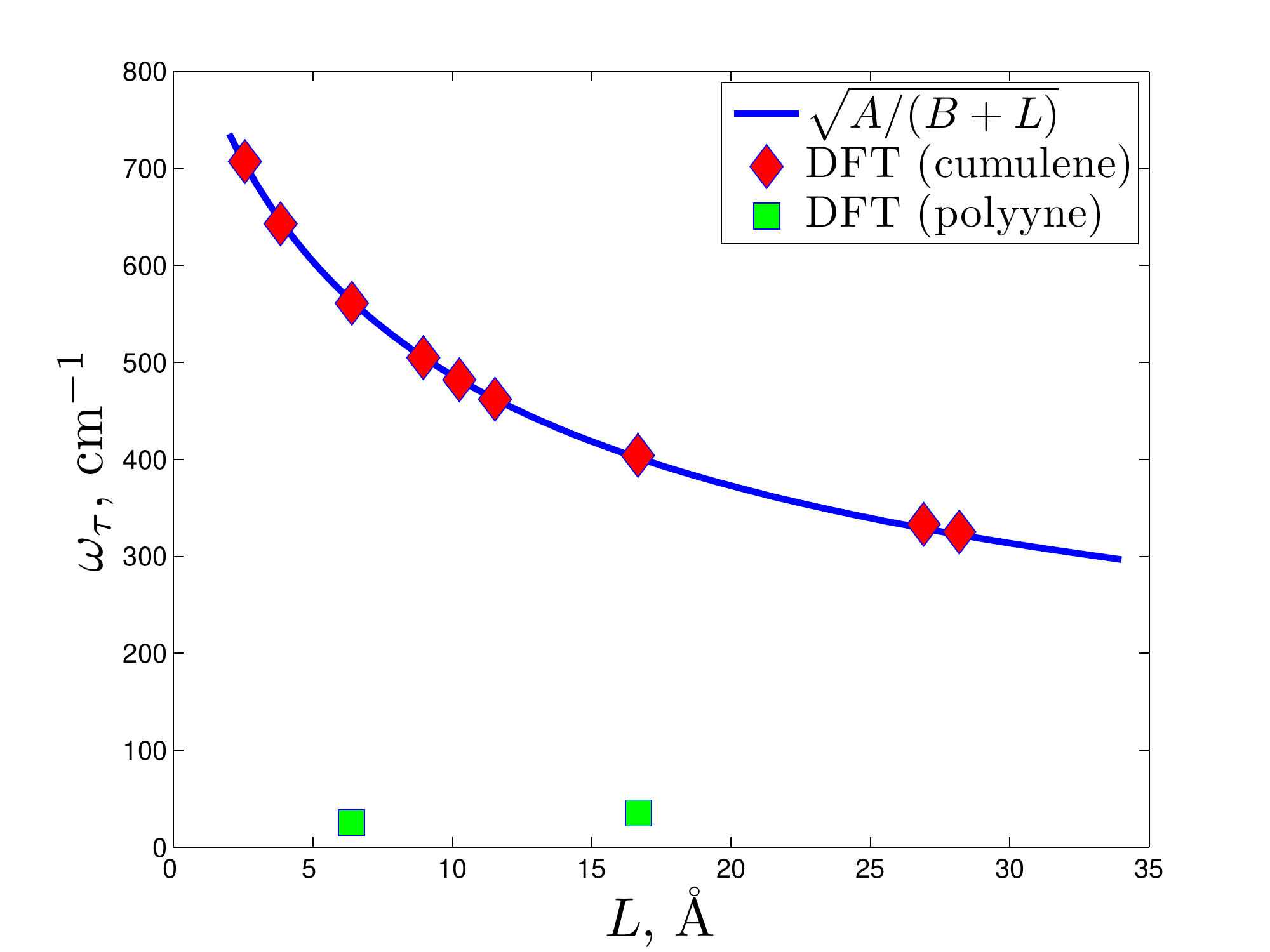}
\caption{\label{Fig:FreqVSLength} Frequency of the torsional mode $\omega_t$ as a function of cumulene molecule length calculated using $A = 3.37$~cm$^{-2}\cdot$\AA, $B = 4.22$~\AA. Atomic moment of inertia $J = 3.43$~u$\cdot$\AA $^{2}$. From Eq.~(\ref{Eq:TorsFreq}) $\kappa = 2.89\cdot 10^6$~cm$^{-2}\cdot$u$\cdot$\AA$^3$. In addition, torsional mode frequency for polyyne H$_3$C$-$(C$\equiv$C$-$)$_m$CH$_3$, $m = 3, 11$, are shown (green squares); one can see that its torsional stiffness is small and does not follow Eq. (\ref{Eq:TorsFreq})}
\end{figure}

The torsional energy has a minimum at constant torsional angle gradient $(\partial\theta/\partial z) = (\Phi_r - \Phi_l)/L$ suggesting that electrons adiabatically follow atomic motion. For the only hydrogen torsional oscillator mode one can assume antisymmetric condition $-\Phi_l = \Phi = \Phi_r$. Then the Euler equation for Lagrangian (\ref{Eq:SpringLagr}) is 
\begin{equation}\label{Eq:HarmOsc}
\frac{d^2\Phi}{d t^2} = -\frac{4\kappa}{JL}\Phi
\end{equation}   
This equation describes the harmonic oscillator with the frequency defined as  
\begin{equation}\label{Eq:TorsFreq}
\omega_{\tau}^2 = \frac{4\kappa}{J}\frac{1}{L}
\end{equation}
Using the same DFT calculation, from harmonic vibrational analysis one can find $\omega_{\tau}$ of H$_2$C$_n$H$_2$ for different $n$. In Figure~\ref{Fig:FreqVSLength} we represented the related frequency $\omega_{\tau}$ for $n = 5, 6, 8, 10, 12, 16, 24, 25$. Since the choice of length $L$ includes some arbitrariness (our choice is illustrated in Figure~\ref{Fig:JeVSz}), the correct fit should include some length parameter $B \sim a$, so that $\omega_{\tau}^2 = A/(B + L)$. Using optimum fitting analysis we found $B = 4.22~\angstrom$ and the torsional stiffness is given by $\kappa = AJ/4\simeq 2.89\cdot 10^6$~cm$^{-2}\cdot$u$\cdot\angstrom^3\simeq 10.6$~eV$\cdot\angstrom$ (u stands for the atomic mass unit), while atomic moment of inertia $J$ is defined by end groups only and does not depend on $n$.  
These estimates were used to evaluate the speed of \textit{torsitons}. Below we analyze the effect of zero-point atomic vibrations on their spectrum. 


\section{\label{Sec:Gap}Effect of zero-point atomic vibrations}
In our description of the electronic torsional mode we implicitly used Born-Oppenheimer approximation, considering electronic motion in an axially symmetric field of motionless nuclei, positioned along the z-axis. This axial symmetry is reflected by the symmetry of the Lagrangian in Eq. (\ref{Eq:RodLagr}) with respect to the change of the function $\theta(z)$ by arbitrarily constant.  

In reality, the nuclei participate in zero-point vibrations in the ground state, which does not possess an axial symmetry because this ground state is adjusted to the electronic ground state where this symmetry is broken (see Figure~\ref{Fig:C5H4Twist}). Indeed, to find this ground state, one needs to consider interacting nuclei in the field of electronic cloud with already calculated anisotropic electronic density. 

Thus, the potential energy depends on the angle $\theta$ even in the absence of torsion and the energy minimum is realized at some angle $\theta_{0}$ which we can set to zero. The potential energy can be expanded over the small displacement from this minimum as $\alpha\theta^2/2$. This term incorporated to the Lagrangian in Eq. (\ref{Eq:RodLagr}) as $-\alpha\theta^2/(2L)$ leads to the gap in the spectrum of torsional waves.  Correcting Eq. (\ref{Eq:FastModeMotion}) by $-\alpha\theta/(j_eL)$ term in the right hand side, we obtain a new dispersion relation
\begin{equation}\label{Eq:DispersionGap}
\omega^2(q) = \frac{\kappa}{j_e}q^2 + \frac{\alpha}{j_e L}
\end{equation}
Since $\omega(0)\neq 0$ the mode is not exactly acoustic due to the gap $\Delta\omega = \sqrt{\alpha/j_eL}$. 

To estimate the parameter $\alpha$  consider the change of classical energy $\delta E(\theta) = \left<\Matr{H}(\theta) - \Matr{H}(0)\right>_g$, where $\Matr{H}$ is the atomic chain quantum Hamiltonian and $\left<\dots\right>_g$ is an average over the ground state of carbon atoms considering their zero-point vibrations. 

All the normal modes of carbon atoms in the molecule, which don't include motion of hydrogen atoms with respect to the adjacent carbon atoms, can be either longitudinal or transverse. For $D_{2d}$ symmetry point group with coordinate system defined above there are only two possible second order invariants: $z^2$ and $x^2 + y^2$ \cite{wilson2012molecular}, so the transverse modes of the harmonic Hamiltonian of the atomic chain are expected to be double-degenerate and the corresponding eigenfunctions possess axial symmetry. Practically, this degeneracy is observed in DFT-calculated IR spectra of H$_2$C$_{n}$H$_2$ for odd $n$, while for even $n$ all  energy levels are split, because such molecules belong to $D_{2h}$ symmetry group. Indeed, the splitting is entirely an effect of sides, because in even $n$ molecules the side CH$_2$ groups lie in the same plane (while in odd $n$ they are orthogonal), so X-Z and Y-Z plane become distinguishable, while for an infinite chain this effect would disappear.

The break of axial symmetry takes place in the third order anharmonic interaction. To express potential energy in normal modes representation introduce notations $u_{x_i}$ and $u_{y_i}$ for transverse modes with energy $\hbar\omega_{x_i} = \hbar\omega_{y_i} = \hbar\omega_i$ and $u_{z_k}$ for $k-th$ longitudinal mode. Thus the third order anharmonic energy is expressed by 
\begin{multline}\label{Eq:V3}
\Matr{V}_3 = \sum_{ijk}\Big\{V_{x_iy_jz}u_{x_i}u_{y_j} +\\
 \frac{V_{x_ix_jz_k}}{2}\Bigl(u_{x_i}u_{x_j} + u_{y_i}u_{y_j}\Bigr)\Big\}u_{z_k}
\end{multline}  
where it is assumed that $V_{x_ix_jz} = V_{y_iy_jz}$ for any $i, j$.

With $\Matr{V}_3$  as a perturbation, a meaningful correction to the ground state in the first order of perturbation theory is given by
\begin{equation}\label{Eq:GroundState}
\left|\psi_0\right> = \left|0\right>  - \frac{1}{\hbar\sqrt{8}}\sum_{ijk}\frac{V_{x_iy_jz_k}}{\omega_i + \omega_j + \omega_{z_k}}\left|1^x_i 1^y_j 1^z_k\right>
\end{equation}
The rotation of electronic cloud about the z-axis by an angle $\theta$ changes the energy in diabatic approximation by
\begin{equation}\label{Eq:dH}
\delta\Matr{H}(\theta) = -2\sum_{ijk}V_{x_iy_jz_k}u_{x_i}u_{y_j}u_{z_k}\sin^2\theta
\end{equation}
Assuming for the small displacement from the minimum $\sin\theta\simeq\theta$, one can find $\delta E(\theta) = \left<\psi_0\right|\delta\Matr{H}(\theta)\left|\psi_0\right> = \alpha\theta^2/2$, with $\alpha$ given by 
\begin{equation}\label{Eq:alpha}
\alpha = \frac{1}{2\hbar}\sum_{k}\Big\{\sum_{i}\frac{V^2_{x_iy_iz_k}}{2\omega_{x_i} + \omega_{z_k}} + \sum_{i<j}\frac{2V^2_{x_iy_jz_k}}{\omega_{x_i} +  \omega_{y_j} + \omega_{z_k}}\Big\}
\end{equation} 

Using anharmonic frequency analysis of H$_2$C$_5$H$_2$ \cite{g09, barone2005anharmonic} we calculated third-order anharmonicity constants. To exclude effect of hydrogen atoms we considered the only transverse and longitudinal  normal modes with the nearest integer of reduced mass greater or equal 2 atomic units. Applying Eq. (\ref{Eq:alpha}) we found $\alpha = 3.2$~cm$^{-1}$, so that the energy gap can be estimated using Eq. (\ref{Eq:DispersionGap}) as  $\Delta\varepsilon\simeq 130$~cm$^{-1}$. 

To answer a question how crucial is the described effect for the acoustic mode, one can find the length $L$ of a cumulene chain where this energy becomes comparable to the minimum \textit{torsiton} energy, which can be estimated as
\begin{equation}\label{Eq:MinEnergy}
\hbar\omega_{min} \simeq \hbar cq_{min} = \hbar c\frac{\pi}{L}
\end{equation}
Thus the length required to make the gap value of the same order as $\hbar\omega_{min}$ is $L_*\simeq \hbar\pi c/\Delta\varepsilon \simeq 128$~nm, that is much larger than the real molecule length \cite{januszewski2014synthesis}. 

Axial symmetry can be violated also by the forth-order anharmonic interaction, however its contribution into the energy gap does not change qualitatively the presented estimate.
\section{\label{Sec:Experiment}Experiment suggestions}

\begin{figure}
\centering
\includegraphics[scale = .2]{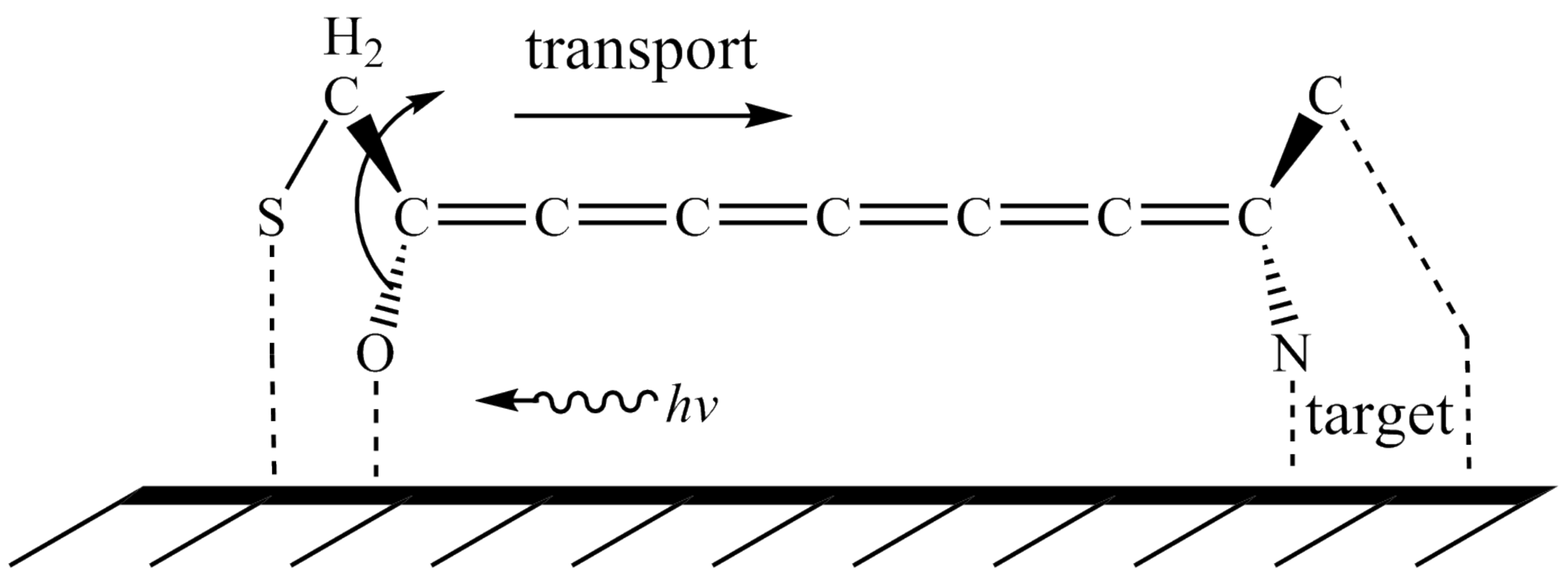}
\caption{\label{Fig:ExperimentScheme} A schematic  experiment set up on remote chemistry initiation. The compound features two surface-anchored end-groups connected by a cumulene chain. Laser initiated bond breaking at the "initiation" (left) end-group can result in generation of a strong torque at the chain which will propagate as a wave-packet along the chain and can result in bond breaking at another end group, the "target".}
\end{figure}

As shown above, the electronic torsional mode features an unprecedented speed of 1000~km/s = 1~nm/fs and can transfer energy up to 10~eV, which is comparable to the energies of the strongest chemical bonds (C=C, N$\equiv$N, etc.). Such high transferred energy brings an opportunity of performing chemistry at distances, including chemical bond breaking reactions. Figure~\ref{Fig:ExperimentScheme} shows a schematic of the compound suitable for the proof of principle experiment on remote chemistry initiation. The compound features two surface-anchored end-groups connected by a cumulene chain. Laser initiated bond breaking at the "initiation" (left) end-group can result in generation of a strong torque at the chain which will propagate as a wave-packet along the chain and can result in bond breaking at another end group, the "target". The energy released by the initiation end-group can be tuned by selecting convenient functional groups. Spectroscopic observation of the transported energy can aim at detecting the formation of the products at the target or detection of the excess energy at the target. In the latter case a longer cumulene chain is required as for the chain length of 50 carbon atoms the transport time is only ca. 5~fs. Compounds with such long chains have been synthesized for polyynes \cite{chalifoux2010synthesis} and we hope that this should be possible for cumulenes as well.
\section{\label{Sec:Conclusions}Conclusion}
In this article we considered electronic torsional waves in cumulene chains (\textit{torsitons}) which are torsional sound waves of entirely electronic nature. We evaluated the speed of \textit{torsiton} propagation  as high as 1000~km/s. Single \textit{torsiton} can carry energy up to 10~eV. 
Similar waves should exist in other atomic chain with anisotropic bonds including recently discovered transition metal linear chains. While the largest band energy computed for cumulenes at 10~eV, the computations neglected electronic excitation, which will likely be contributing at such high energies. It will be interesting to see how the ground electronic state \textit{torsitons} are perturbed by electronic excitations at higher \textit{torsiton} band energies and how the quasi-particles of two types, \textit{torsitons} and excitons, interact. Nevertheless, the presented band calculations are expected to be free of electronic excitation effects at smaller energies. Importantly, the transport speed supported by the lower half of the \textit{torsiton} band is similar to that of the full band (with small corrections due to the \textit{torsiton}-vibron coupling). 
 

\begin{acknowledgments}
We thank Marina Matherne for suggestion of cumulene chain to consider, Dmitry Polyanski for the discussion of bond breaking reaction at distance, and Noa Marom for suggestions on electronic density calculation. This work was supported by National Science Foundation (CHE-1462075) program.
\end{acknowledgments}

\bibliography{BibliographyForCarbyneShort}
\end{document}